\documentclass{article}
\usepackage[utf8]{inputenc}

\usepackage{graphicx}
\graphicspath{ {./Figures/} }

\usepackage{multicol}
\usepackage{authblk}
\usepackage{textcomp}
\usepackage{xcolor}
\usepackage{biblatex}
\title{Topic Modeling in Density Functional Theory on Citations of Condensed Matter Electronic Structure Packages}

\author[1, 3]{Marie Dumaz \thanks{Corresponding Author: mcd0029@mix.wvu.edu}  }
\author[2]{Camila Romero-Boh\'orquez}
\author[3]{Donald Adjeroh}
\author[1]{Aldo H. Romero}

\affil[1]{
Department of Physics and Astronomy,
West Virginia University,
Morgantown, WV 26506-6315, USA
}

\affil[2]{
Neuroscience Department,
West Virginia University,
Morgantown, WV 26506-6315, USA
}

\affil[3]{
Lane Department of Computer Science and Electrical Engineering,
West Virginia University,
Morgantown, WV 26506-6315, USA
}

\date{}

\begin{document}

\maketitle

\begin{abstract}
  With an increasing number of new scientific papers being released, it becomes harder for researchers to be aware of recent articles in their field of study. Accurately classifying papers is a first step in the direction of personalized catering and easy access to research of interest. The field of Density Functional Theory (DFT) in particular is a good example of a methodology used in very different studies, and interconnected disciplines, which has a very strong community publishing many research articles. We devise a new unsupervised method for classifying publications, based on topic modeling, and use a DFT-related selection of documents as a use case. We first create topics from word analysis and clustering of the abstracts from the publications, then attribute each publication/paper to a topic based on word similarity. We then make interesting observations by analyzing connections between the topics and publishers, journals, country or year of publication. The proposed approach is general, and can be applied to analyze publication and citation trends in other areas of study, beyond the field of Density Function Theory. 
\end{abstract}

\section{Introduction}

For many years, scientific documents in basic sciences were classified using the Physics and Astronomy Classification Scheme (PACS) method, which maps a series of codes with keywords to cover the main topics of the publication. This is usually done by the author, and provides relative freedom for the actual classification. Though this is still one of the methods used by several journals, this step is transitioning into a more focused type of topic classification\cite{smith2019pacs}. Basically, this change is driven by the fact that the institute that introduced this approach (the American Institute of Physics, AIP) has decided to discontinue its support, primarily  due to the huge cost of administration and paper handling during manuscript submission.
Therefore, a more dynamical and flexible method is necessary. 
A method were different topics capture the continuous change in the scientific development and scientific interest and are able to map reviewers and even journal in this arena. In this respect, the use of topic modeling from machine learning is gaining a lot of attention~\cite{vayansky2020review, barde2017overview}. In this paper, we propose an unsupervised learning technique that focuses on finding text correlations and thus classify documents according to a small set of topics. These topics are found through an unsupervised machine learning technique that creates clusters of meaningful words in the document selection, by finding similarity between them in a latent space. 
Though the idea of topic modeling has been used in the past to  classify research topics~\cite{paul2009topic, zhao2014topic}, in this work,  we start from a well-defined database, and apply this methodology on publications related to the use of electronic structure packages. 

While a general method to categorize topics in basic sciences would be ideal, in this paper, we have decided to focus in a particular research field, where we can have access to a very well controlled database~\cite{dumaz2021authorship}. Density Functional Theory (DFT) is the most used methodology to characterize materials at the atomic scale~\cite{kohn1965self, hohenberg1964inhomogeneous}. It falls into the so called electronic structure methods and it has the benefit that a large number of computational codes and software employ this method, which have created communities around them, supporting the research efforts. We base our document selection on research papers citing those computational code libraries, in order to reduce our scope to only DFT-related papers. 

Some efforts to analyze DFT-related papers from different subfields have been made, but they classify the articles based on the PACS numbers \cite{aleta2019explore, chinazzi2019mapping}. However, our dataset  downloaded from the Web of Science database, does not keep records of the PACS numbers and only retains keywords. Moreover, as DFT is a methodology with many different applications, in fields other than physics, not all papers follow the PACS structure. This shows again, the necessity for a new classification system, based purely on the content of the documents. Our approach is not only able to classify documents on content but also can be used to find novel topics and relationships between topics which can be used to identify new and emerging research fields.

\section{Methodology}

The research papers and their abstracts were obtained using the same methodology as in previous work \cite{dumaz2021authorship}, through the Web of Science database (WoS), but expanded to include 2020 articles.

Once we extracted the abstracts from the WoS database, we performed several pre-processing steps to capture only the most meaningful words out of the text. First, we deleted certain groups of words, such as ``left arrow" or ``vertical bar," that are only used to create mathematical representations. For the same reason, we removed all characters between two dollar signs, as this is a standard command to make equations, which would only confuse the algorithm. There is one exception to this rule if an underscore is found between the two dollar signs, as that pattern is used for chemical formulas that are essential and meaningful notions in sub-fields of DFT. We also remove publishing and copyright tags, DOI numbers, and all non-alphanumeric characters. 

Then, the leftover sentences get tokenized, which means the text is broken down into units, creating a list of tokens from the text. From those lists, we take out all stopwords, or words that are so common they do not help differentiate different sub-fields. We used the standard NLTK \cite{bird2009natural} list of stopwords and added others based on our own experience with the data after rounds of trial and feedback. For example, ``theory" and ``method" are not specific to some areas of physics like magnetic properties or gas absorption. We also remove all single letters and digits, often used in mathematical expressions or to quantify a value and don't hold meaning intrinsically. We then create bigrams and trigrams, groups of two or three tokens that often appear one after the other and will now be linked as one word by an underscore. Examples would be ``ab initio" and ``density functional theory." Finally, the tokens are lemmatized to reduce words to their root, and the resulting list of tokens is filtered to only keep the ones that are in at least two documents and in less than 20\% of the papers. From this final list are created a dictionary and a Bag of Words (BoW) which are the input to our algorithm.

Latent Dirichlet Allocation (LDA) \cite{blei2003latent} is a topic modeling algorithm which uses Bayesian statistical methods to generate text, from documents passed in as input in the BoW. This means that LDA does not take into account the order of words. In fact, an LDA model creates a set of topics where each topic is a probability distribution over a corpus of words drawn from the distribution $\beta_k \sim \mbox{Dirichlet}(\eta)$, and each document is a probability distribution over topics drawn from $\theta_k \sim \mbox{Dirichlet}(\alpha)$.

\begin{figure}[!htb]
\begin{center}
    \includegraphics[width=0.5\textwidth]{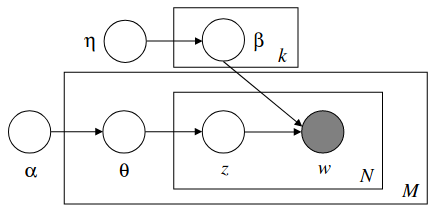}
\end{center}
\caption{Graphical representation of a LDA model \cite{blei2003latent}}
\label{fig:lda}
\end{figure}

Here, we define $M$ as the total number of documents, $N$ as the total number of words in a document $m$, and $k$ as the number of topics. As shown in Figure \ref{fig:lda}, LDA spans over three layers. The parameters $\alpha$ and $\eta$ are corpus-level parameters, which means that they are sampled only once per corpus. $\theta$ denote document-level variables, sampled once per document. Finally, variables $z_n$ and $w_n$ are at word-level: they are sampled once for each word in each document. $w_n$ is also the only observable variable, picked from the bag of words given as input. 

For topic modeling, we implemented a Latent Dirichlet Allocation (LDA) algorithm with the Python library Gensim \cite{rehurek_lrec} that already has a handy set of functions to create, evaluate and visualize LDA models. We used two bag of words (BOWs): one with all documents in our dataset from 1990 to 2019, and one with only documents published in 2020. The 2020 corpus was used as a test set. 

We created 16 models with a different number of topics ($k$), from 5 to 80, while keeping all other parameters null. For this analysis, we chose to tune the parameters left for 15, 25, 35 and 45 topics. All models were evaluated with the $C_v$ coherence score \cite{roder2015exploring}, which is the coherence score with the strongest correlation to human rankings. We then ran 480 models to tune hyperparameters including $\alpha$ and $\eta$, for each of the 5 different values of $k$. For each value of $k$, the model that gives the highest $C_v$ score is used through the rest of this article. 

To speed up training, we run the parameter tuning on several cores. We should note that the stochastic optimization process used in online LDA includes the randomization of data partition, which makes it impossible to ensure exact reproducibility if coupled with parallelization.

To discover trends and interests, we assign topics to our publications bibliometric attributes. We assign topics by considering the probabilities returned by the LDA model as weight. We create a matrix $A$ where each row represents a document, and each column represents a topic. The elements of this matrix are the probabilities for a given topic to be assigned to a given document. 

This matrix is particularly useful for the analyses made of topic distributions over countries, journals and years. This way, when computing the number of publications per topic per country, for example, we can use those weights to perform a more precise analysis. For these analyses, we create a matrix where each row is a topic and each column is a country. If a document was created in the USA for example, then each topic would add their respective weight for this document in the ``USA" column. These analyses are also all normalized, which means each value in the matrix is divided by the sum of all values in the $A$ matrix, which is equal to the total number of documents.

\section{Results}
Here, we describe our dataset, experiments and results. 

To create a dataset of research papers focused on DFT, we used the graphical interface ``Web of Science'' (WOS) and downloaded all articles citing some of the most commonly used computational packages to study crystalline systems. We acquired all papers from the first citation up to 2019 for our training set; and all citations in the 2020 year for our test set. \cite{dumaz2021authorship}

\subsection{Distribution of documents and topics}

\begin{figure}[!htb]
\begin{center}
    \includegraphics[width=0.49\textwidth]{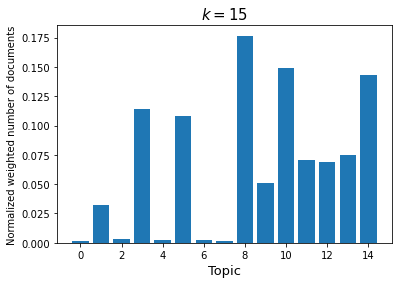}
    \includegraphics[width=0.49\textwidth]{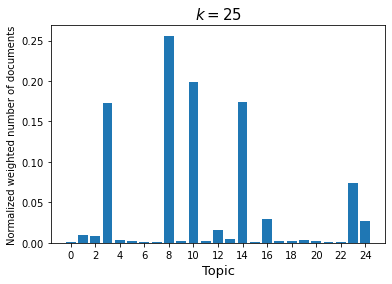} \\
    \includegraphics[width=0.49\textwidth]{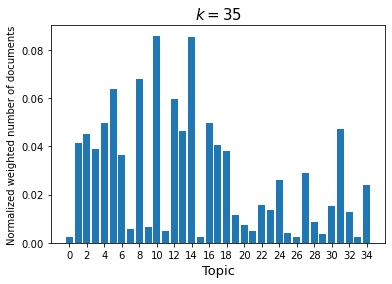}
    \includegraphics[width=0.49\textwidth]{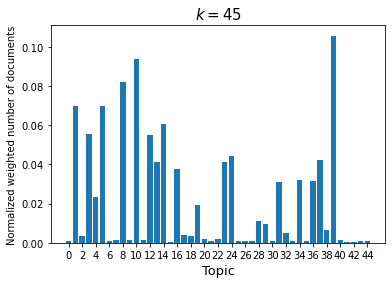}
\end{center}
\caption{Document distribution over topics (for publications from 1990-2019)} 
\label{fig:topic_dist}
\end{figure}

Figure \ref{fig:topic_dist} shows the distribution of documents in our dataset over topics. This means that it displays how many documents contain each topic. Note, we use the probabilities of a topic appearing in a document as weight, as described in the Methodology section. We add up the weights for all documents in each topic, and divide the results by the total number of documents. The figure gives us an idea of how well distributed the topics identified from each model is, and what topic is more common within the documents.

We can see that the model with 25 topics is extremely unbalanced, with only 4 topics accounting for up to 80\% of the documents. Next, the LDA models for $k$ = 15 and $k$ = 45 seem to be somewhat well distributed, as no topics is contained in more than 18\% and 11\% of the documents. This can easily be deduced via the y-axis of the histograms, that are also a quick way to get a general idea of a distribution: if the y-axis range is larger, it means that at least one topic takes a bigger part of the documents, which means the distribution is less uniform. Similarly, we can study the range of the attributed normalized number of documents. The model with 25 topics has the bigger range at 0.255, followed by 15 topics at 0.175 and 45 topics at 0.105. The smallest range, and hence the model that has a more uniform distribution is for $k$ = 35 with 0.086. Indeed, to get 90\% of the documents, it takes the first 8 topics (or 53\% of the topics) with the highest proportion for $k$ = 15, 24\% of the topics for $k$ = 25, 57\% for $k$ = 35 and 37\% for $k$ = 45. Moreover, the model with 35 topics is also the one that has the highest minimum percentage of documents.

The diversity in the models for $k$ = 35 and $k$ = 45 also seems to imply that 15 topics is not enough for our corpus. In fact, we find evidence for topics specialization that further confirm the benefits of a higher number of topics. For example, in the $k$ = 15 model, Topic 3 contains the following words: ``formation", ``activity", ``hydrogen", ``metal", ``reaction", ``mechanism", ``process", ``catalyst", ``oxidation", and ``li". We can map this topic to two different topics in the $k$ = 35 model: it shares the words ``activity", ``metal", ``reaction", ``catalyst", ``oxidation" with Topic 6 and the words ``reaction", ``formation", ``mechanism", ``process" with Topic 27. We can distinctly see that it splits into two categories, one is more focused on specific processes and applications (Topic 6) and one is more general (Topic 27).

We can conclude from these analyses that the most balanced model is for $k$ = 35. For these two reasons, we will mostly focus on this particular LDA model and its results in the rest of this section. 

\begin{figure}[!htb]
\begin{center}
    \includegraphics[width=0.88\textwidth]{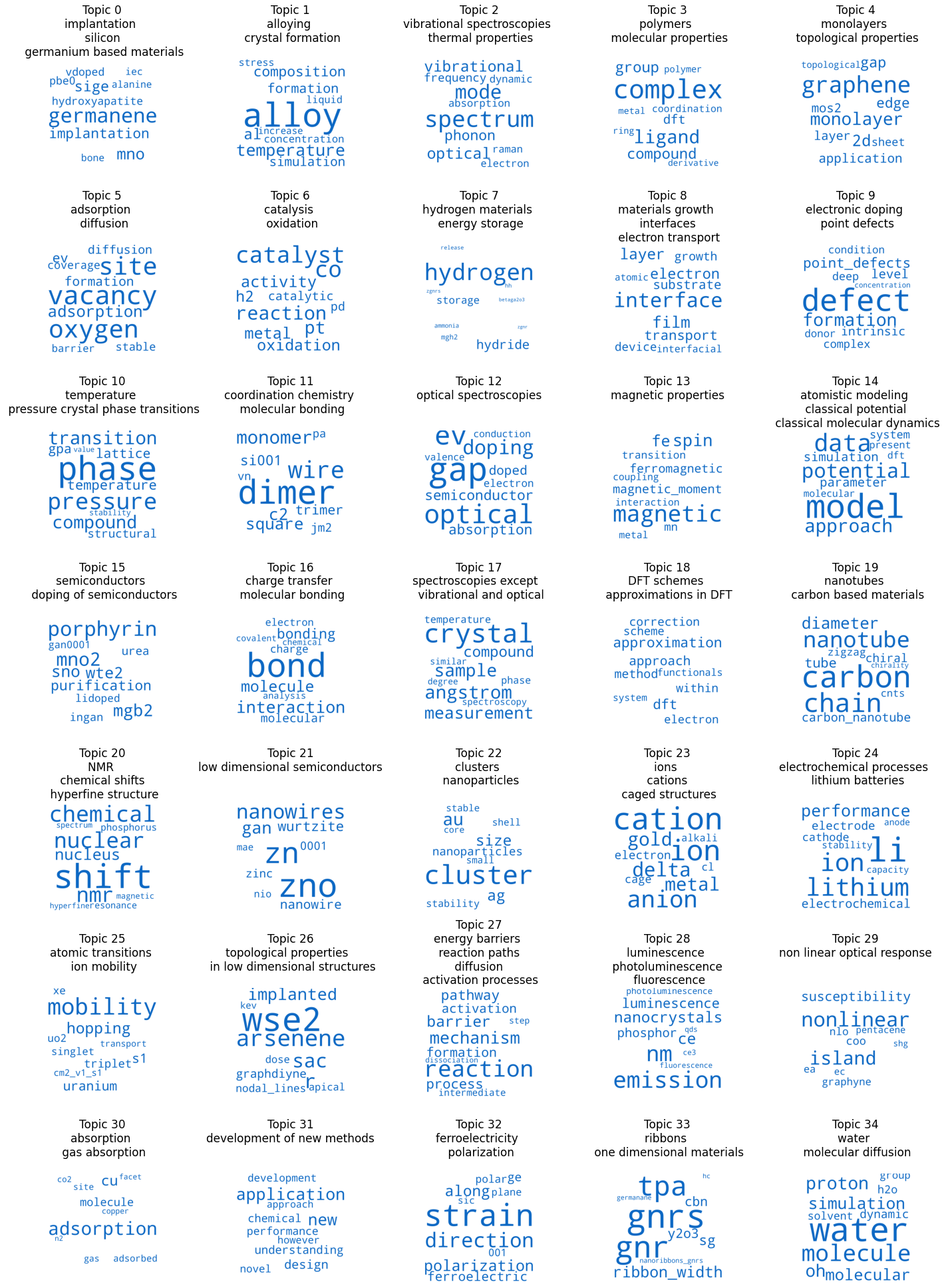}
\end{center}
\caption{Top 10 words making up each topic in the LDA model for $k$ = 35}
\label{fig:wordcloud35}
\end{figure}

\subsection{Closer look at the Topics}
Figure \ref{fig:wordcloud35} shows the word clouds for the 35 topics in our LDA model. Each word cloud represents the 10 words with the highest probability in the topic. The size of each word is proportional to their probability within their topics. Bigger words are terms with a higher probability and a bigger impact in the topic. We also added topic names to give a quick idea of what each word cloud represents. Even though some topics overlap, most topics are rather independent and clearly defined. For example, in Topic 13, we can find words relating to magnetic problems, while Topic 7 was more about hydrogen and hydride. Hydrogen is not a magnetic atom so those two subjects are not expected to be clustered together, and indeed, they were not: they are quite different. Topic 1 on alloy, however, can be related to magnetism as some alloy metals are magnetic and yet, it is once again put into a separate group (Topic 13), which shows how good the LDA model is at detecting different word relationships. Topic 14 is another good example of terms that could apply to many different fields with words such as ``data", ``simulation", ``model" or ``approach", but are associated to each other in a unique topic.

\begin{figure}[!htb]
\begin{center}
    \includegraphics[width=0.99\textwidth]{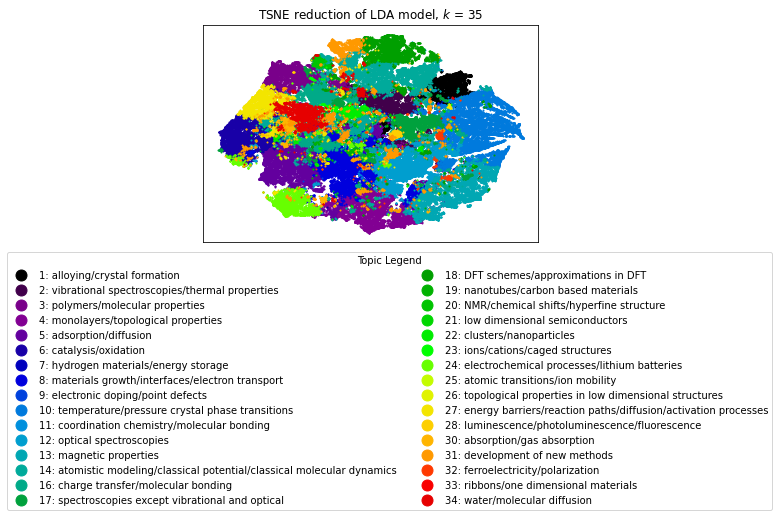}
\end{center}
\caption{T-SNE reduction for the LDA model for $k$ = 35}
\label{fig:tsne}
\end{figure}

We can also analyze the topics of our model by visualizing them with the t-SNE reduction method \cite{hinton2002r}. Figure \ref{fig:tsne} displays the data reduction to 2-dimensions, where each point is a document, positioned depending on their similarity with other documents, and each color is their dominant topic. Note that some topics are not represented as they are not a dominant topic in any document, this is the case for topics 0 and 29 for example. Since the topics are well distributed over the documents, we can see a lot of topics being represented, instead of a few overshadowing the entire map. We can also notice that some topics are more dispersed, such as the subject of new methods (e.g., Topic 31). That is expected as researchers are developing new processes and data simulations in many different fields. However, other topics are clustered, and clearly defined like the first topic on alloying and crystal formation. This figure also helps visualizing the similarity between topics. The closer two topics are, the more interconnected they are. For example, Topic 27 (energy barriers/reaction paths/diffusion/activation processes) is close to Topic 6 (catalysis/oxidation) and Topic 3 (polymers/molecular properties) but is on the opposite side of Topics 1 (alloying/crystal formation) and 13 (magnetic properties). Indeed, molecular properties and activation processes like catalysis are chemistry related while magnetism and alloying are more often studied in physics.

\begin{figure}[!htb]
\begin{center}
    \includegraphics[width=0.49\textwidth]{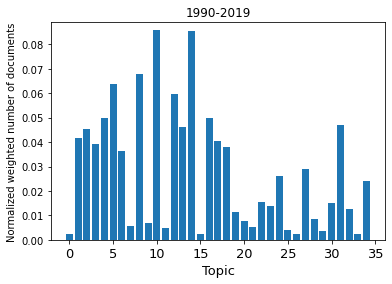}
    \includegraphics[width=0.49\textwidth]{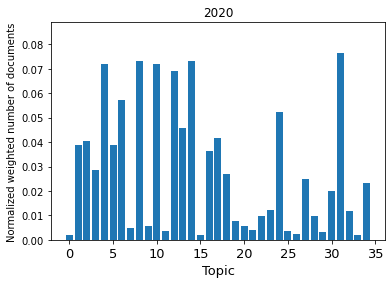}
\end{center}
\caption{Document distribution over topics, for $k$=35 topics. Left: 1990-2019 corpus. Right: 2020 corpus}
\label{fig:topic_dist2020}
\end{figure}

\section{Temporal evolution of topics}
Figure \ref{fig:topic_dist2020} compares the distribution of documents over topics, for the 1990-2019 corpus on the left (that was already shown in Figure \ref{fig:topic_dist}), and for the 2020 corpus on the right. The two plots are similar and both display a good distribution of the documents. While Topic 10 (temperature/pressure crystal phase transitions) and Topic 14 (atomistic modelling/classical potential/classical molecular dynamics) were the most common subjects up to 2019, a number of other topics caught up to them in 2020, including Topic 31 (development of new methods) which now has the highest number of assigned documents. This could also mean that while Topic 8 (materials growth/interfaces/electron transport) and Toipc 10  were very important up to 2019, and still are, Topic 4 (monolayers/topological properties), Topic 24 (electrochemical processes/lithium batteries) and Topic 31 (development of new methods) are gaining more momentum recently. The similarity between the two plots also indicate that our model is generalizing correctly and can efficiently attribute topics to new documents.

\begin{figure}[!htb]
\begin{center}
    \includegraphics[width=0.99\textwidth]{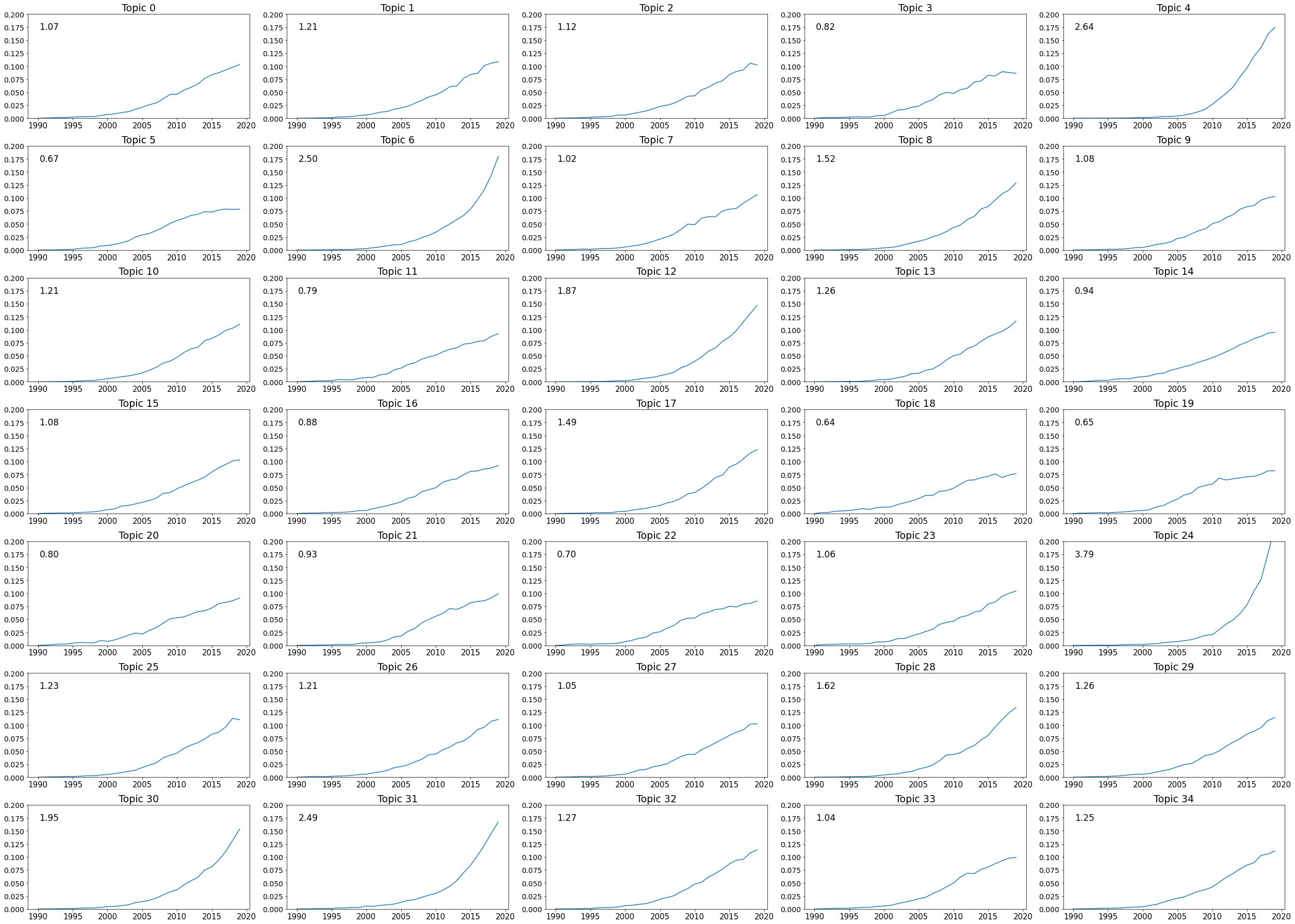}
\end{center}
\caption{Normalized and weighted number of documents over the years, per topic for $k$ = 35. Number inside each plot (top left) denotes the best fitting power factor.}
\label{fig:topic_year}
\end{figure}

We computed, for each topic independently, the normalized and weighted number of documents per year. Every year, there is an exponential amount of DFT-related articles being published, and consequently, almost all topics also grew exponentially. To compare them, we calculated the power law factor of each curve since 2000. Even though some topics' evolution are better fitted with an exponential curve, the power law function yields the highest mean $r^2$ over all topics. The results are shown in Figure \ref{fig:topic_year}, where the number in each top left corner of the subplots are the power factor best fitting for each topic. We chose to only consider publications after 2000 as this is the period in which DFT software really started to to get globally used.

A total of nine topics have a power factor lower than 1 and while they still grew in importance since 2000, a few of them are starting to stabilize and even loose some interest from the DFT community. For example, Topic 18 on DFT schemes and Topic 19 on nanotubes, which have the lowest factors recorded, have not gained any new momentum in the recent years and Topic 18 also observed a decrease of publications in 2017 for the first time ever. On the contrary, Topics 24 related to electrochemical processes and lithium batteries, as well as Topic 4 (monolayers/topological properties), Topic 6 (catalysis/oxidation) and Topic 31 (development of new methods) are growing at a significantly high rate and seem to be among the most important subjects in the field of DFT today. Other subjects are all developing at intermediate rates, representing the diversity of DFT and how important the methodology is for different topics. 

\begin{figure}[!htb]
\begin{center}
    \includegraphics[width=0.49\textwidth]{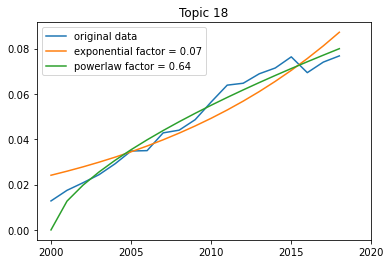}
    \includegraphics[width=0.49\textwidth]{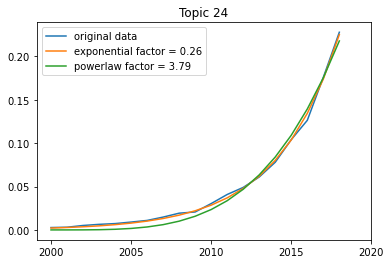}
\end{center}
\caption{Comparison of the two topic with the lowest and highest growth, fit to the best exponential curve and powerlaw curve. Left: Topic 18 with lowest power law factor. Right: Topic 24 with highest power law factor.}
\label{fig:years_extremes}
\end{figure}

We report in Figure \ref{fig:years_extremes} the two topics with the smallest and highest growth out of the 35 in our model. We fit a power law and an exponential curve to both of the curves and reported the factors in the top left corner. Topic 24 (electrochemical processes/lithium batteries) is an example of a growth better fitted with an exponential function, getting a $r^2$ of 0.998, even though a power law function also get a high $r^2$ score of 0.988. On top of the important difference in growth, we can also notice how ``bumpy" the evolution of Topic 18 (DFT schemes) is. The many irregularities, compared to the smooth curve of Topic 24 might also be a sign of waning (or at least, inconsistent) interest. 

\begin{figure}[!htb]
\begin{center}
    \includegraphics[width=0.95\textwidth]{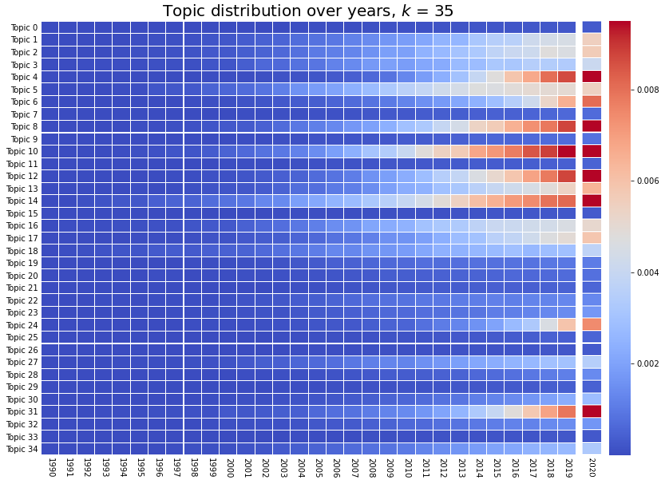}
\end{center}
\caption{Normalized and weighted number of documents over the years, per topic, for $k$ = 35}
\label{fig:PYheatmap}
\end{figure}

Figure \ref{fig:PYheatmap} also shows the growth of topics over the years, in the form of a heat map. We notice that no important increase happened before 2000, which confirms our choice to study the exponential factors only after 2000. Moreover, most topics only start gaining momentum after at least 2005. It is interesting to note that the three topics with the highest exponential factors (Topic 4, 6 and 24) grew extremely fast in a few years, which is the reason for the high factor. Topic 24 especially only started to differentiate itself from the others just in 2018. It is also interesting how out of those three topics, two of them (Topics  6 and 24) are not part of the topics with the highest proportion of documents. However, given their high power factor, we can only predict that they will become more important in the next few years. Similarly, by looking at the trajectory of some subjects such as Topics 1 or 17, we can confidently forecast that the interest in those subfields will grow in the coming years. The 2020 column in this figure seems to agree with those predictions as Topics 1, 4, 6, 17 and 24 all gained in proportion during the last year. On the other hand, we can also easily determine which topics are stabilizing and are loosing momentum. For example, Topic 3 (polymers/molecular properties) and Topic 18 (DFT schemes/approximations in DFT) have had interest from the scientific community for at least a decade and yet, are not growing as fast as others. 

It is interesting to compare the evolution of Topics 14 and Topic 31, as they are related to each other. Topic 14 contains words such as ``data", ``model" and ``simulation" which represent the community's effort to incorporate code, machine learning and new computing methods into their work. This topic has been growing steadily over the years, as it becomes a bigger part of today's experiments. Secondly, Topic 31 relating to the improvement and optimization of code and processes, only started gaining momentum in 2014, about 10 years after Topic 14. However, its growth was faster, which places both of the topics at the top of the most published subjects in 2020 in DFT. We can also mention the evolution of Topic 4 related to graphene and monolayers. It started to grow a few years after 2010, which is the year that the Nobel Prize in physics was awarded to Andre Geim and Kostya Novoselov for their work regarding the two-dimensional material graphene.

\begin{figure}[!htb]
\begin{center}
    \includegraphics[width=0.95\textwidth]{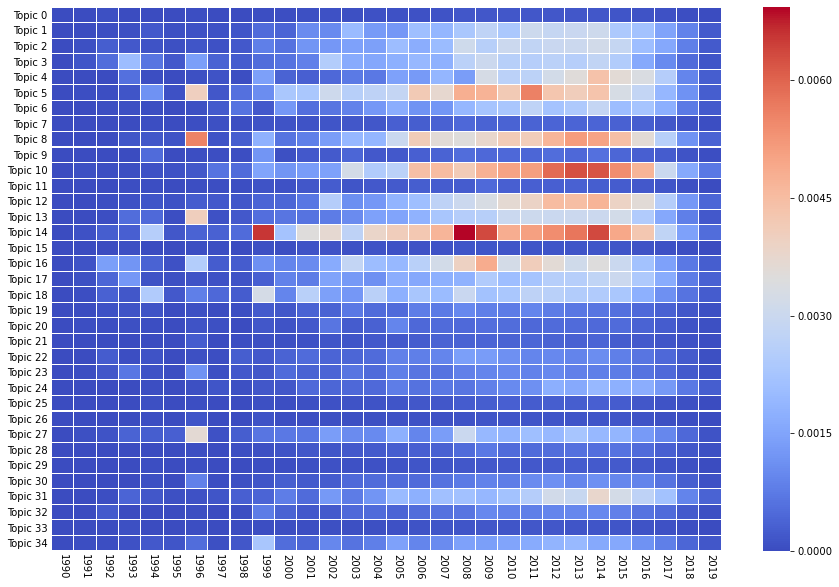}
\end{center}
\caption{Normalized and weighted number of citations over the years, per topic, for $k$ = 35}
\label{fig:cit_PYheatmap}
\end{figure}

\section{Trends in citations}
Looking at the number of citations per year, per topic in Figure \ref{fig:cit_PYheatmap}, we notice the same trends as noted in the previous figure, only pushed a few years back. Indeed, papers published today are citing older research articles so a raise in publications in 2019 means a raise in citations of older papers, typically at least 3 years prior. This gap is surely due to the research process of reading the literature, putting together a new idea and hypothesis, getting data and experiences and finally drawing conclusions and then publishing. 

Hence, looking at the trends of citations over the years might give us more insight and more predictive value than only the number of published papers. For topics that are only starting to attract more interest, like Topics 6 and 24, we can trace back the beginning of the trend to the late 2000s or early 2010s whereas topics that have been in place for longer such as Topics 8, 10 or 14 can be attributed to discoveries in the late 1990s.

\begin{figure}[!htb]
\begin{center}
    \includegraphics[width=0.49\textwidth]{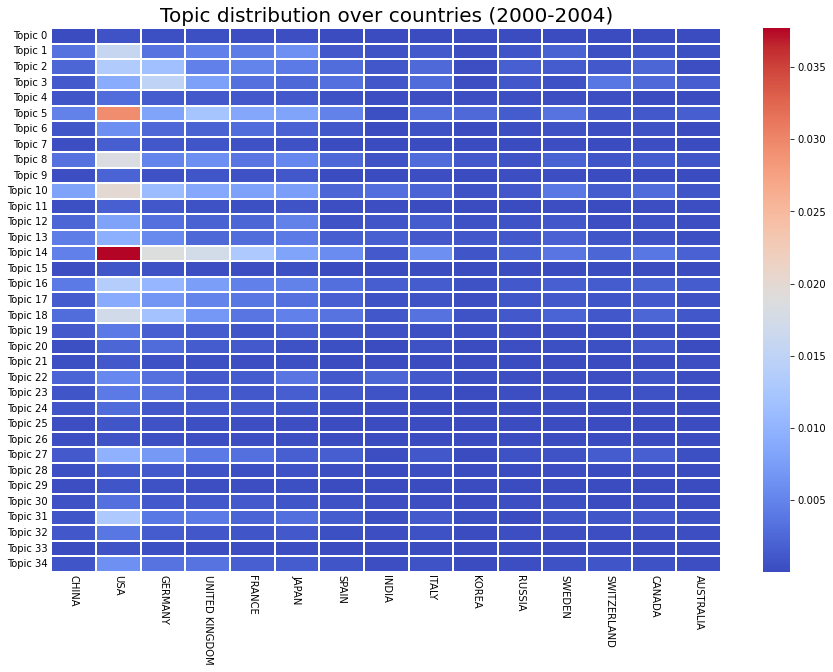}
    \includegraphics[width=0.49\textwidth]{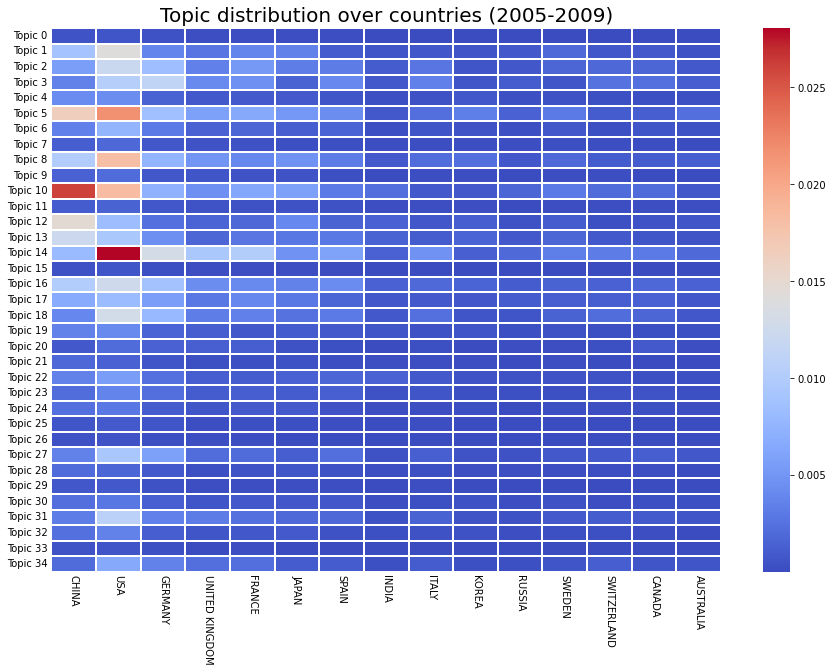} \\
    \includegraphics[width=0.49\textwidth]{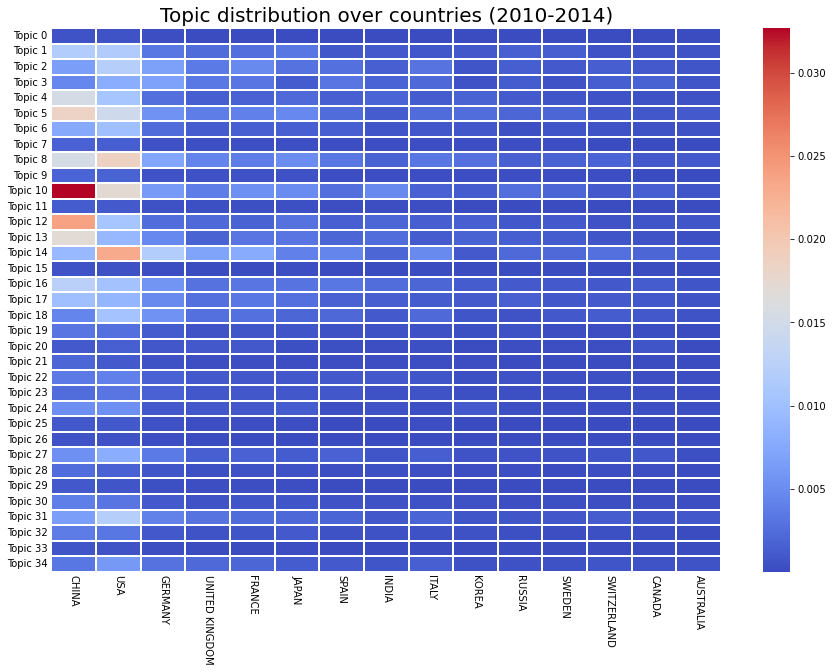}
    \includegraphics[width=0.49\textwidth]{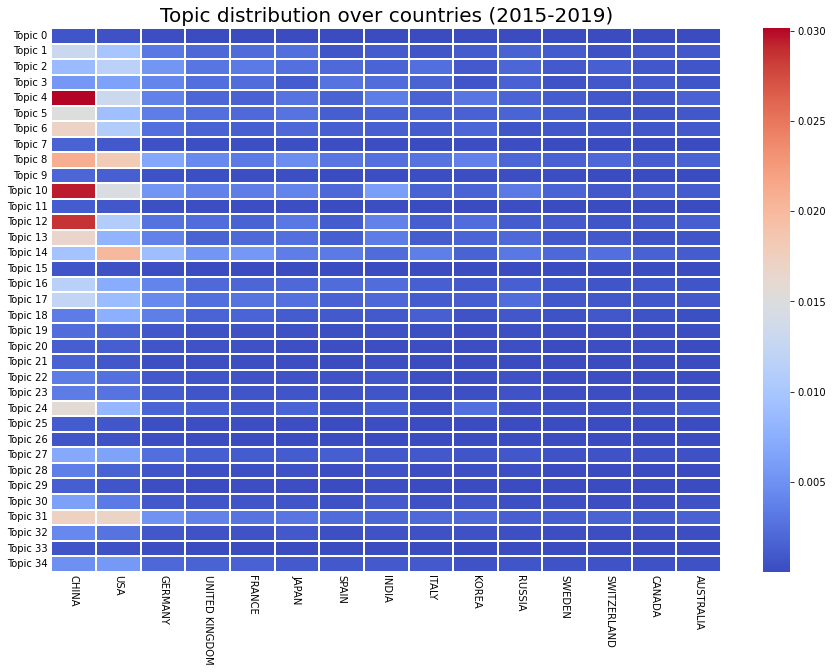}
\end{center}
\caption{Normalized and weighted number of documents per country (only top 15 shown), per topic, in different time periods, for $k$ = 35}
\label{fig:COheatmap}
\end{figure}

\subsection{Spatial evolution of topics}
We also studied the spatial evolution of topics in DFT over the years, especially, by countries. To do so, we identify countries using the affiliations of all authors. However, we only count a country once for each paper.

Figure \ref{fig:COheatmap} divides the heat maps of topics per country over four different time periods. Before 2005, the USA were dominating the field of DFT and published the highest number of publications in a larger number of topics. Then, the proportions slowly shifted to China, which now generates more articles, and diversified its interests. China started to get involved in DFT after 2005 through the subject of crystal phase transitions (Topic 10). Pressure crystal phase transitions is the main strength of DFT and one of the most active subfields, which explains why it would be the first topic to attract interest. Topic 10 stayed an important part of work in China until now, as it was still one of the main published subjects by authors from China between 2015 and 2019, along with Topics 4 and 12. Nowadays, the USA still contributes to many different topics in DFT, and grew interest in Topic 14, but decreased its number of publications in Topic 5 (adsorption/diffusion).

\begin{figure}[!htb]
\begin{center}
    \includegraphics[width=0.9\textwidth]{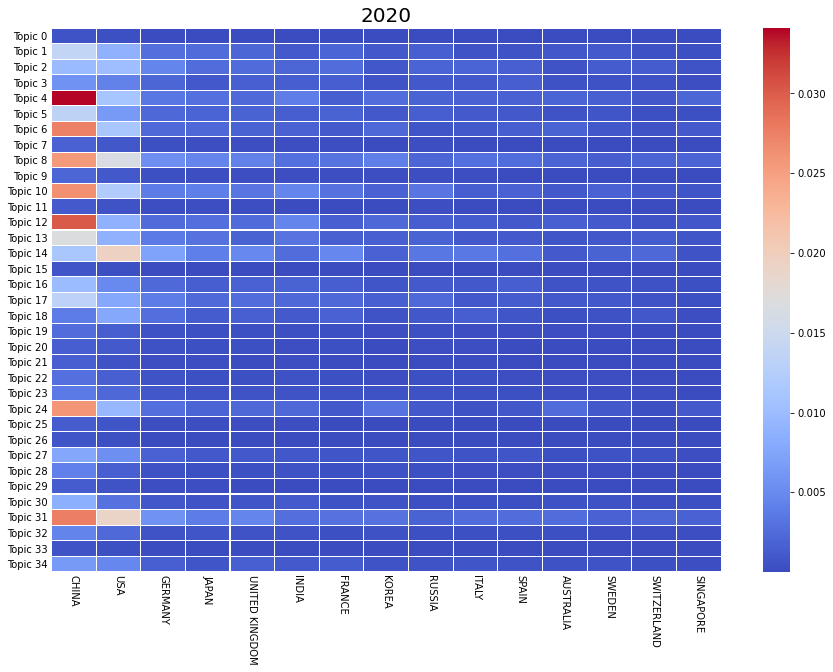}
\end{center}
\caption{Normalized and weighted number of documents per country (only top 15 shown), per topic, for articles published in 2020, for $k$ = 35}
\label{fig:COheatmap_2020}
\end{figure}

We can accurately predict that China will soon dominate the field of DFT as its number of publications will only keep growing. Figure \ref{fig:COheatmap_2020}, representing the normalized distribution of documents per topics over countries, for the year 2020 confirms the influence of China in DFT. Moreover, even though most software counted in this work are developed in Europe, they are the most used in China and the USA. Germany, the third country with the highest number of publications, and country of origin to Turbomole, the sixth most used code in our dataset is barely even competing with China and the USA after 2005. Similarly, the impact of the United Kingdom and France  kept decreasing over the years, even though they were very active in several different topics before 2005.

\section{Topics distribution by journals and publishers}

\begin{figure}[!htb]
\begin{center}
    \includegraphics[width=0.49\textwidth]{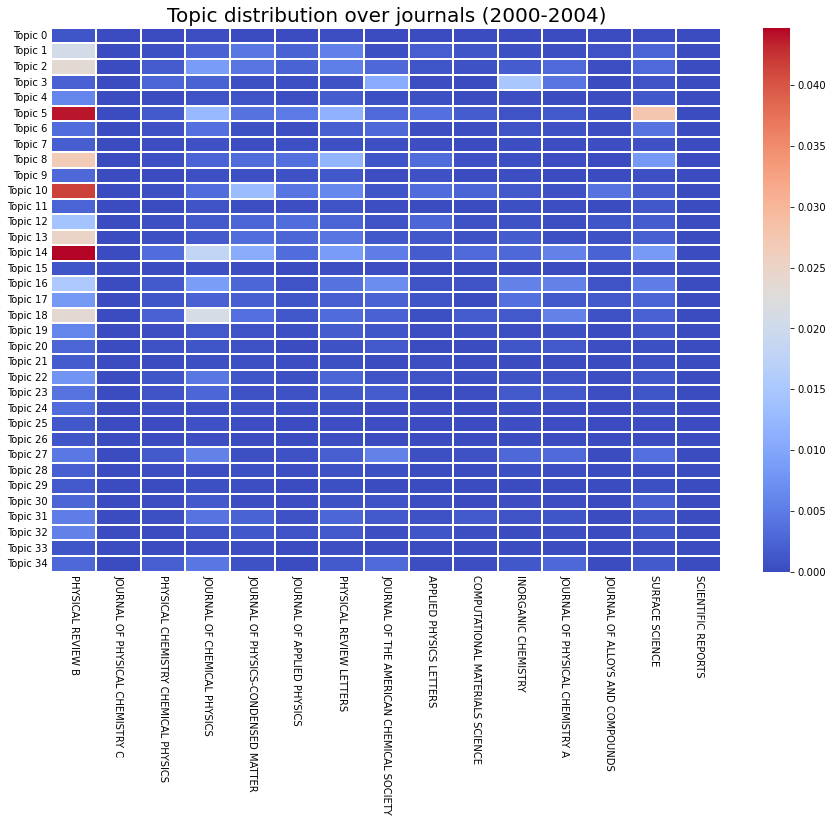}
    \includegraphics[width=0.49\textwidth]{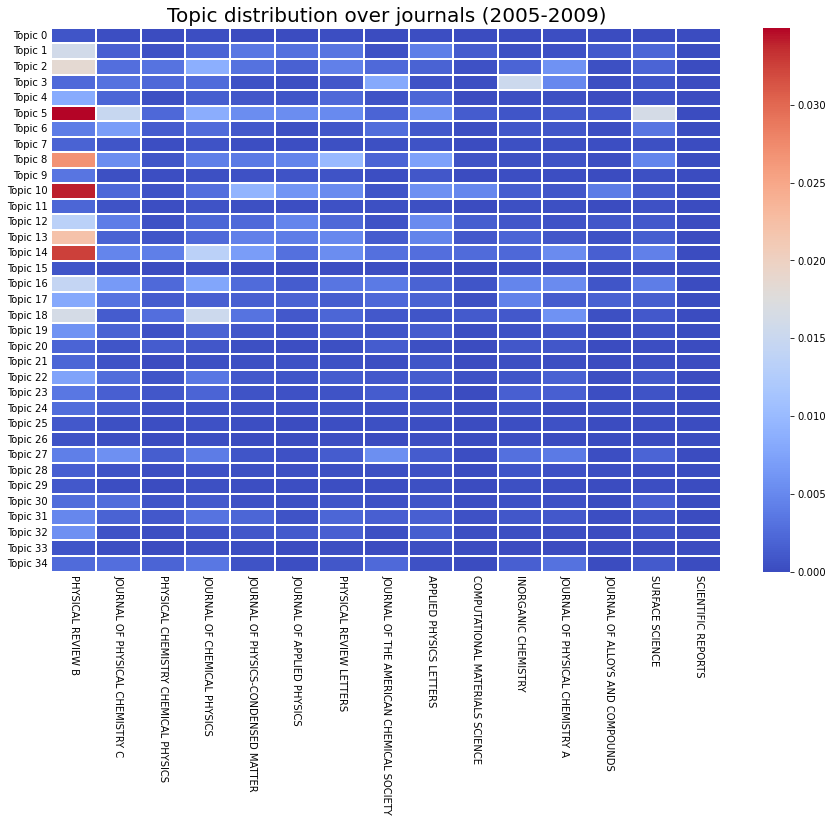} \\
    \includegraphics[width=0.49\textwidth]{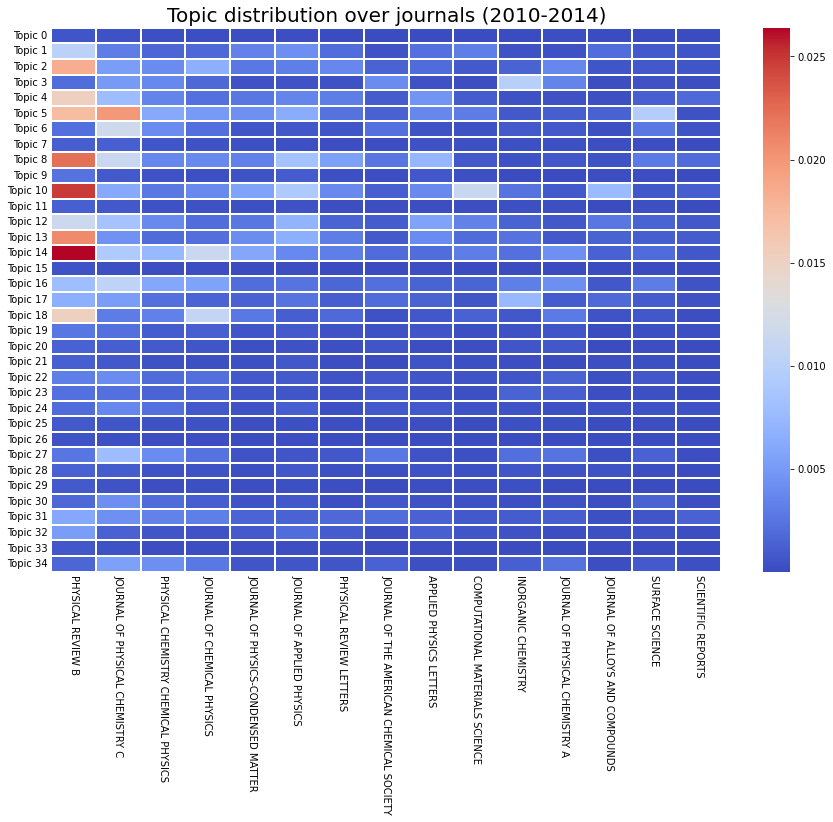}
    \includegraphics[width=0.49\textwidth]{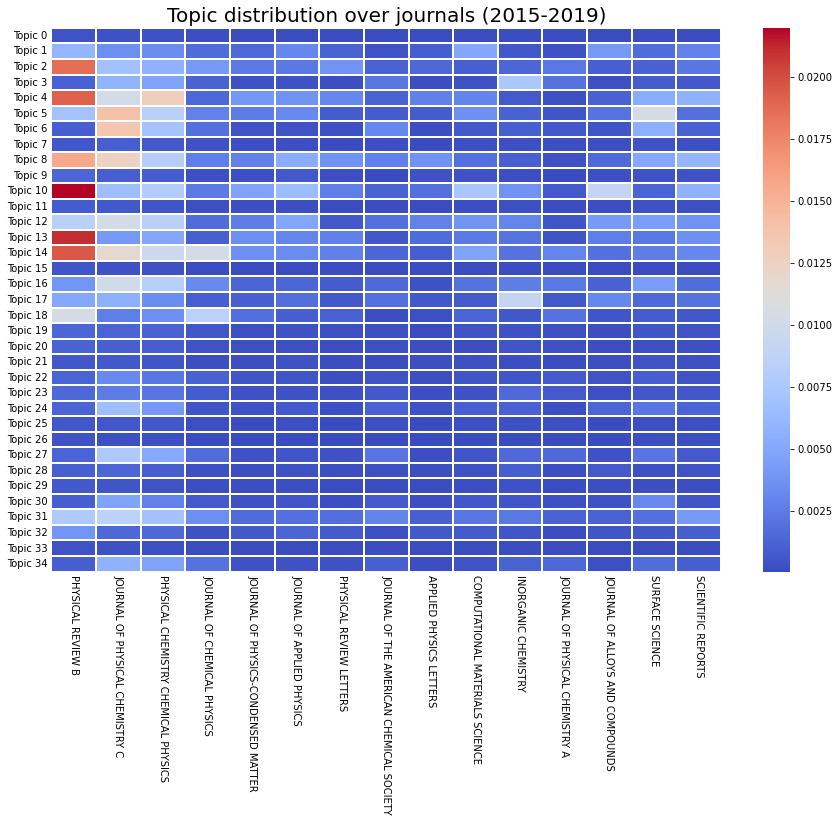}
\end{center}
\caption{Normalized and weighted number of documents per journal (only top 15 shown), per topic, in different time periods, for $k$ = 35}
\label{fig:journal_heatmap_over_time}
\end{figure}

Similarly to figure \ref{fig:COheatmap}, figure \ref{fig:journal_heatmap_over_time}, shows the normalized number of publications (computed with the probabilities as weight) per topic over journals, dividing the papers over four 5-year periods based on their publication year. In the same manner, these plots express the diversification of DFT publications over different journals, while still highlighting the importance of a very small number of prominent members. However, while figure \ref{fig:COheatmap} displays signs of diversification amongst countries as soon as 2005, most papers were still published in Physical Review B until a lot more recently, and no other journal seems to be taking over its number of publications.

The 15 journals displayed have the highest number of publications and are arranged in order from left to right. Even though some journals like Surface Science are specializing in one topic, and sticking with it over time, we can notice how most journals are present in several different subjects. Two key journals, Physical Review B and Journal of Physical Chemistry C are both influential in DFT and were expected to be the largest contributors to the field. Moreover, as DFT is mostly used in physics and chemistry, it is not surprising that those two journals both focused on condensed matter analysis each represent one of the two domains. 

In fact, the three journals in positions 2, 3 and 4 are related to the the chemistry side of DFT. Even though, they have similar targets, they evenly distribute the publications between them. The Journal of Chemical Physics focuses on Topic 14 and 18, Journal of Physical Chemistry C contributes more to Topics 5, 6, 8 and a few others while Physical Chemistry Chemical Physics has a more diverse range of subjects. Hence, it seems that the Journal of Chemical Physics contributes more to methodologies and not material characteristics, judging by the words present in Topics 14 and 18 (modeling/approximations). Even though physics researchers mostly publish their articles in one particular journal, chemists have more options available depending on their interests.

\begin{figure}[!htb]
\begin{center}
    \includegraphics[width=0.9\textwidth]{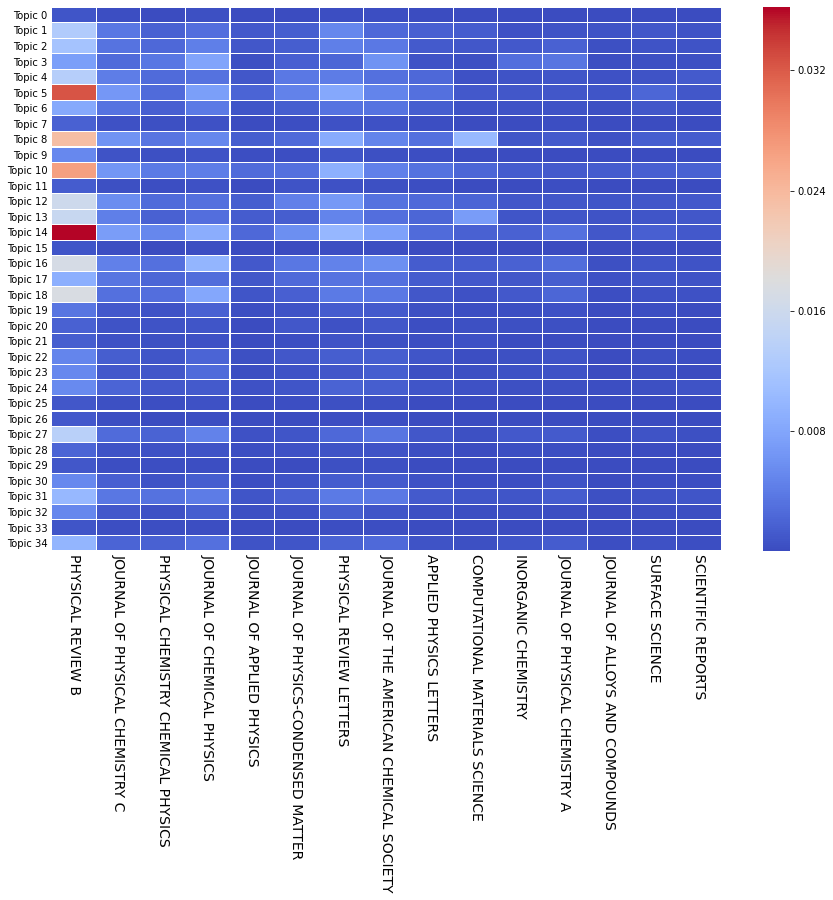}
\end{center}
\caption{Normalized and weighted number of citations per topic and per journal (only top 15 shown), with only open access journals considered, for $k$ = 35}
\label{fig:cit_journal}
\end{figure}

Figure \ref{fig:cit_journal} seem to confirm our deductions from the previous figure. However, even though journals publish papers in a lot of different topics, they seem to be cited for only a few, which might steer their editorial agenda in the future. In fact, the four topics most cited in Physical Review B, are also the four topics most published in the same journal. With such hypothesis we could predict a raise in the number of papers published in Topics 16 and 27, as they are more cited and published, as of 2019. We should note that even tough the values are normalized, a few highly cited articles might skew the results. 

\begin{figure}[!htb]
\begin{center}
    \includegraphics[width=0.9\textwidth]{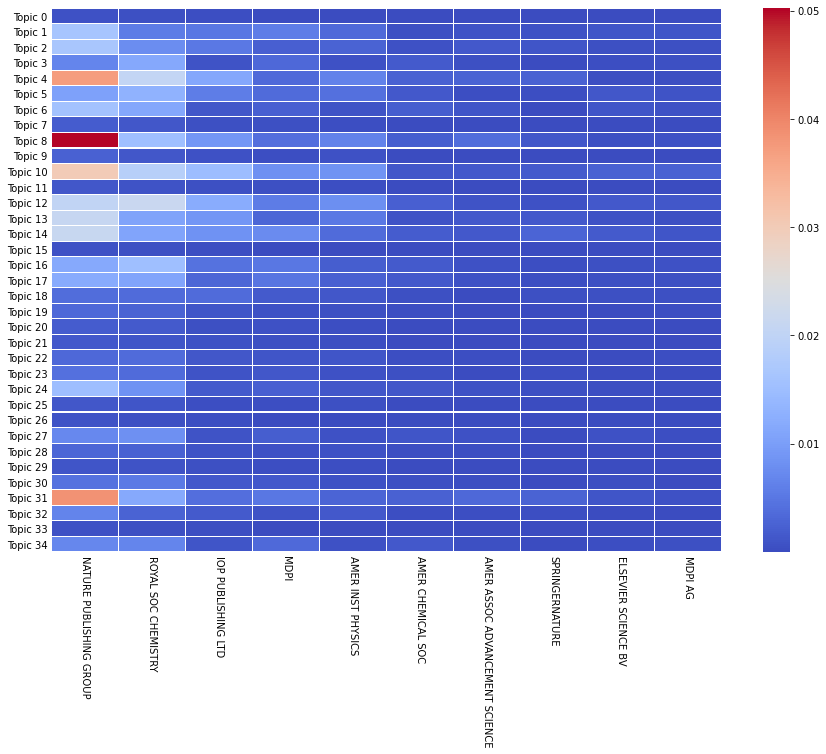}
\end{center}
\caption{Normalized and weighted number of documents per topic and per publisher (only top 10 shown), with only open access journals considered, for $k$ = 35.}
\label{fig:openjournal_heatmap}
\end{figure}

Figure \ref{fig:openjournal_heatmap} is similar to Figure \ref{fig:journal_heatmap_over_time}, but the x-axis shows publishers (and not journals). The other difference is that we created this plot with only articles published in open-access journals. It is interesting to see how the two figures have different interests. For example, open-access journals seem to have a strong interest in Topic 31 while subscription journals in Figure \ref{fig:journal_heatmap_over_time} only published a small amount of their articles relating to this subject. The other three most important topics in open-access journals are Topics 4, 8 and 10 which are also well published in paid journals but to different levels. While Topics 4 and 8 grew in proportion, Topics 10, 13 and 14 all saw their share of documents decrease in open-access publishers, even though they still remain relevant. Moreover, it seems that open-access journals are more focused on a few topics, instead of spreading to many different subjects, like paid journals. We can also notice the rather good correlation with the evolution of topics over time as the most published papers in open-access journals are also the ones that are the most published in 2020 (see Figure \ref{fig:PYheatmap}).

\section{Discussion}

Bibliometrics and topic modeling are powerful tools that can be used to study and evaluate the evolution and impact of any given field. In this paper, we developed several analyses related to the evolution of topics with emphasis on the specific field of DFT. Here, evolution was considered in terms of time (over the years), and space (across countries), and also with respect to journals, publishers, institutions, and subjects based on information  from over 120,000 papers in Density Functional Theory. We find that automatically classifying papers based on their abstract is a promising option for improved, faster, and more objective categorization. In fact, most of the 35 topics resulting from our LDA model were consistent with our document corpus and could be attributed a name.

Even though most packages are developed in Europe, they are mostly used in the USA and China, that are dominating the amount of publications created in the DFT field. While the two countries share interests in a lot of different subjects, American scientists work with many of the most important DFT packages, and researchers in China use mostly VASP, which is a licensed package. This difference might come from the intrinsic set of tools each software has, and the subfields they serve the most. Indeed, China seems to direct most of its work to materials science and applications as they focus on 2D materials (Topic 4), chemical reactions (Topic 6) or batteries (Topic 24) while researchers in the USA published most of their  work in simulations (Topic 14) and new methodologies (Topic 31).

Moreover, two of the most recurrent topics in our analysis, and that receive the most interest from influential journals and countries are related to code development, simulations and modeling. As our work focuses on computational packages, it is not surprising that some papers focus on the methods and processes of software. However, the large amount of interest in the subjects suggests a recent and strong effort in the physics and chemistry fields to develop and optimize code packages and libraries.

The topic modeling done in this article was created using the abstracts of papers and already resulted in coherent and clear topics. Future work could include replicating our methodology on the full-text of all 120,000 papers in our dataset.  Further, the approach used is quite general, and can easily be adapted to apply to other fields of study, beyond DFT.

\section*{Additional Information} 

The authors declare no competing interests.

\section*{Data Availability} The dataset that supports the findings of this study are available in Figshare with the identifier
https://doi.org/10.6084/m9.figshare.12494654.v2.

\printbibliography
    
\vspace{15mm}

\section*{Author Contributions} 

M.D. and C.R.-B. downloaded the data from the Web of Science database and preformed data cleaning and data pre-processing. M.D. executed model training and testing. M.D., D.A. and A.H.R. did the data analysis and interpretation, created the data figures and wrote the main text. All authors reviewed the manuscript.

\vspace{5mm}

\newpage

\section*{Figures}

\vspace{4mm}

\noindent

\textbf{Fig. 1} Graphical representation of a LDA model \cite{blei2003latent}

\vspace{4mm}

\textbf{Fig. 2} Document distribution over topics (for publications from 1990-2019)

\vspace{4mm}

\textbf{Fig. 3} Top 10 words making up each topic in the LDA model for $k$ = 35

\vspace{4mm}

\textbf{Fig. 4} T-SNE reduction for the LDA model for $k$ = 35

\vspace{4mm}

\textbf{Fig. 5} Document distribution over topics, for $k$=35 topics. Left: 1990-2019 corpus. Right: 2020 corpus

\vspace{4mm}

\textbf{Fig. 6} Normalized and weighted number of documents over the years, per topic for $k$ = 35. Number inside each plot (top left) denotes the best fitting power factor.

\vspace{4mm}

\textbf{Fig. 7} Comparison of the two topic with the lowest and highest growth, fit to the best exponential curve and powerlaw curve. Left: Topic 18 with lowest power law factor. Right: Topic 24 with highest power law factor.

\vspace{4mm}

\textbf{Fig. 8} Normalized and weighted number of documents over the years, per topic, for $k$ = 35
\vspace{4mm}

\textbf{Fig. 9} Normalized and weighted number of citations over the years, per topic, for $k$ = 35

\vspace{4mm}

\textbf{Fig. 10} Normalized and weighted number of documents per country (only top 15 shown), per topic, in different time periods, for $k$ = 35

\vspace{4mm}

\textbf{Fig. 11} Normalized and weighted number of documents per country (only top 15 shown), per topic, for articles published in 2020, for $k$ = 35

\vspace{4mm}

\textbf{Fig. 12} Normalized and weighted number of documents per journal (only top 15 shown), per topic, in different time periods, for $k$ = 35

\vspace{4mm}

\textbf{Fig. 13} Normalized and weighted number of citations per topic and per journal (only top 15 shown), with only open access journals considered, for $k$ = 35

\vspace{4mm}

\textbf{Fig. 14} Normalized and weighted number of documents per topic and per publisher (only top 10 shown), with only open access journals considered, for $k$ = 35.

\end{document}